\documentclass{aa}
\usepackage{txfonts}
\usepackage{graphicx}
\usepackage[section]{placeins}

\newcommand{\dScu}{{$\delta$~Scuti }}
\newcommand{\gDor}{{$\gamma$~Doradus }}

\bibpunct{(}{)}{;}{a}{}{,}

\begin{document}

\title{The envelope of the power spectra of over a thousand \\ 
\dScu stars\thanks{Table~A.1 is only available in electronic form at the CDS via anonymous ftp to cdsarc.u-strasbg.fr (130.79.128.5) or via http://cdsweb.u-strasbg.fr/cgi-bin/qcat?J/A+A/}}
\subtitle{The $\bar{T}_{\rm eff}-\nu_{\rm max}$ scaling relation}

\author{S. {Barcel\'o Forteza}\inst{\ref{ins0},\ref{ins1},\ref{ins1.1}} \and T. {Roca Cort\'es}\inst{\ref{ins1},\ref{ins1.1}} \and R.~A. Garc\'ia\inst{\ref{ins3}} }

\institute{Departamento de Astrof\'isica, Centro de Astrobiolog\'ia (CSIC-INTA), ESAC, 28691, Villanueva de la Ca\~nada, Madrid, Spain \label{ins0}
\and
Instituto de Astrof\'isica de Canarias, 38200 La Laguna, Tenerife, Spain \label{ins1}
\and
Departamento de Astrof\'isica, Universidad de La Laguna, 38206 La Laguna, Tenerife, Spain \label{ins1.1}
\and
Laboratoire AIM, CEA/DRF – CNRS – Univ. Paris Diderot – IRFU/SAp, Centre de Saclay, 91191 Gif-sur-Yvette Cedex, France\label{ins3}
}

\date{Received 21 August 2017; Accepted 16 February 2018}

\abstract{\textit{CoRoT} and \textit{Kepler} high-precision photometric data allowed the detection and characterization of the oscillation parameters in stars other than the Sun. Moreover, thanks to the scaling relations, it is possible to estimate masses and radii for thousands of solar-type oscillating stars. Recently, a $\Delta\nu - \rho$ relation has been found for \dScu stars. Now, analysing several hundreds of this kind of stars observed with \textit{CoRoT} and \textit{Kepler}, we present an empiric relation between their frequency at maximum power of their oscillation spectra and their effective temperature. Such a relation can be explained with the help of the $\kappa$-mechanism and the observed dispersion of the residuals is compatible with they being caused by the gravity-darkening effect.}


\maketitle

\section{Introduction}
\label{s:intro}

Our knowledge of the stellar properties is increasing as asteroseismic techniques are being developed and applied to the huge amount of data available. Ground-based telescopes and networks allowed a preliminary study of the characteristics of the oscillation modes in stars. Some of these networks, such as the Delta Scuti Network \citep[DSN,][]{Breger1998} and the Stellar Photometric International network \citep[STEPHI,][]{Michel1995}, studied \dScu stars with higher spectral resolution, thus improving the observing windows, and allowing the identification of several high amplitude modes. However, the golden age for Asteroseismology came with the launch of the space telescopes such as \textit{CoRoT} \citep{Baglin2006} and \textit{Kepler} \citep{Borucki2010}. These space missions revealed the pulsation pattern of solar-type oscillators (including red giants), thereby turning this new domain into a flourishing field.\\

Moreover, \dScu stars power spectra started to be better measured and better understood \citep[e.g.][]{Poretti2009}. This kind of stars is representative of the whole intermediate-mass domain just above the domain of solar-like pulsators in terms of effective temperature \citep[from $\sim$6000 to $\sim$9000 K;][]{Uytterhoeven2011} and mass \citep[from 1.5 to 2.5 $M_{\odot}$;][]{Breger2000}. They show from slow to fast rotation rates, as it is common in stars in this mass range, and also more massive stars \citep{Royer2007}.\\

The oscillations of these stars are excited by $\kappa$-mechanism \citep{Chevalier1971}. However, turbulent pressure can be of importance too in certain peculiar cases \citep{Antoci2014} or in \dScu - \gDor hybrids \citep{Xiong2016}. The pulsation pattern of \dScu stars also shows regularities as has been supposed in several theoretical studies \citep[e.g.][]{Pasek2012}. These regularities include the large separation whose value is related to the mean density of the star \citep[e.g.][]{Suarez2014}. Such a scaling relation is quite similar to the scaling relation found in solar-type oscillators \citep{Kjeldsen1995} with a small deviation (from 11 to 21\%) due to the stellar deformation. \citet{GarciaHernandez2015} confirm this relation by taking into account several \dScu stars observed in eclipsing binary systems, which suggests that the relation is independent of the rotation rate:
\begin{equation}
\frac{\rho}{\rho_{\odot}} = \left(1.55^{+1.07}_{-0.68}  \right) \left( \frac{\Delta \nu}{\Delta \nu_{\odot}} \right)^{2.035 \pm 0.095} \, .
\label{e:lsep_agh}
\end{equation} 
In addition, the higher and lower limits of the pulsation pattern might be used to characterize the stars \citep{Michel2017}.\\

In this paper, we analyse a large sample of 1063 \dScu stars, observed using CoRoT and Kepler, to characterize and study the envelope of their oscillation power spectra and look for new scaling relations. In Section~\ref{s:dSBF}, the study of the envelope of the spectrum is explained. The different results we obtained are presented in Section~\ref{s:400dScu}, including the linear relation between the frequency at maximum power, $\nu_{\rm max}$, and the effective temperature, $T_{\rm eff}$, of \dScu stars. In Sect.~\ref{s:gdeex}, we further explain the observed dispersion of the measured temperature. In the final section we draw our conclusions.\\

\section{Data and Methodology}
\label{s:dSBF}

A large sample of \dScu type stars is needed for a proper statistical study of their characteristics. Therefore, we have selected 1055 A and F stars with oscillations in the typical frequency range of \dScu type stars observed with \textit{Kepler}. The \textit{Kepler} Long Cadence data (LC) are sampled every $\sim$30 min and its noise level is around 20 ppm \citep{Koch2010}. We also use the data of 8 \dScu stars observed with the \textit{CoRoT} sismo-channel with a much shorter cadence (32s) and a substantially higher photometric precision \citep[between 0.6 to 4 ppm, see][]{Auvergne2009}.\\

Using our own methodology \citep[\dScu Basics Finder;][]{BarceloForteza2017}, we obtained the overall parameters of the oscillation modes and also characterized their power-spectral structure. Our analyses interpolate and analyse the light curve of each star with an iterative three-stage method \citep[see][]{BarceloForteza2015}: The first two stages allow us to interpolate the light curve using the information of the subtracted peaks. This interpolation minimizes the effect of gaps and considerably improves the background noise, thereby avoiding spurious effects \citep{Garcia2014}. The last stage produces more accurate and precise results in terms of the parameters of the modes (frequencies, amplitudes, and phases). Furthermore, it is reasonably fast, thus making it appropriate for the study of a large sample of \dScu stars. Owing to the low Nyquist frequency of the LC data (283 $\mu$Hz), we also made a superNyquist analysis \citep{Murphy2013} up to 1132 $\mu$Hz. This range includes all typical frequencies for \dScu stars \citep{Aerts2010}.\\

To characterize the power spectral structure, we take into account the energy of the signal for each peak $i$:
\begin{equation}
 E_{i} =\frac{RMS_{i}-RMS_{i+1}}{RMS_{0}} \, ,
\label{e:esignal}
\end{equation}
where $RMS_{i}$ is the root mean square of the residual signal after the extraction of highest $i$ peaks, $i=0$ for the original signal). \\

A number of parameters defining the envelope can be obtained; for instance, the number of peaks that form the envelope ($N_{\rm env}$); those that accomplish $E_{i} \gtrsim 0.1 \%$ within the typical \dScu frequency regime, from 60 to 930 $\mu$Hz \citep{Aerts2010}. The level of 0.1\% is chosen since the number of peaks per energy from 0.01 to 0.1\% is an order of magnitude higher than those between 0.1 to 1\%. The energy of all these peaks ($E_{\rm env}$) compares the energy of the envelope with the energy of the entire signal. The mean amplitude of the envelope,
\begin{equation}
 A_{\rm env} = \frac{\sum^{N_{\rm env}}_{i=0} A_{i} }{N_{\rm env}} \, ,
\label{e:Aenv}
\end{equation}
where $A_{i}$ is the amplitude of each peak; and the frequency at maximum power,
\begin{equation}
 \nu_{\rm max} = \frac{\sum^{N_{\rm env}}_{i=0} A_{i} \nu_{i}}{\sum^{N_{\rm env}}_{i=0} A_{i}} \, ,
\label{e:nup}
\end{equation}
where $\nu_{i}$ is the frequency of each peak. This definition is based on the weighted mean frequency used by \cite{Kallinger2010b} to measure the asymmetry of the power excess for red giants. We also measured the asymmetry of the \dScu stars' envelope using
\begin{equation}
 \alpha = \frac{\nu_{\rm max} - \frac{\nu_{h}+\nu_{l}}{2} }{\delta \nu_{\rm env}} \, ,
\label{e:asym}
\end{equation}
where $\nu_{l}$, and $\nu_{h}$ are the lowest, and highest frequency peaks, respectively, and $\delta \nu_{\rm env}$ is its width defined as $\delta \nu_{\rm env} = \nu_{h}-\nu_{l} $. We note that, $|\alpha| = 0.5$ is the maximum value for an asymmetric envelope.\\

\section{Results}
\label{s:400dScu}

Once their power spectra are obtained, we can classify the studied stars in $\delta$~Scuti, $\gamma$~Doradus, and hybrid stars, as explained in \citet{Uytterhoeven2011}. We find that 700 of the 1055 stars ($\sim$66\%) are \dScu stars, 225 stars ($\sim$21\%) are $\delta$~Sct/$\gamma$~Dor hybrids, 116 stars ($\sim$11\%) are $\gamma$~Dor/$\delta$~Sct hybrids, and 14 stars ($\sim$2\%) are \gDor stars or other kinds of pulsators. We take into account those stars without significant pulsation in the \gDor regime since hybrid stars have a higher convective efficiency \citep{Uytterhoeven2011}. This is of importance to accomplish all of our assumptions (see Section~\ref{ss:fast}). Only three stars are considered HADS candidates (see Sect.~\ref{ss:hads}). The star sample is listed in the online-only material (Table~A.1).\\

\begin{figure*}
\centering
\includegraphics[width=0.495\textwidth]{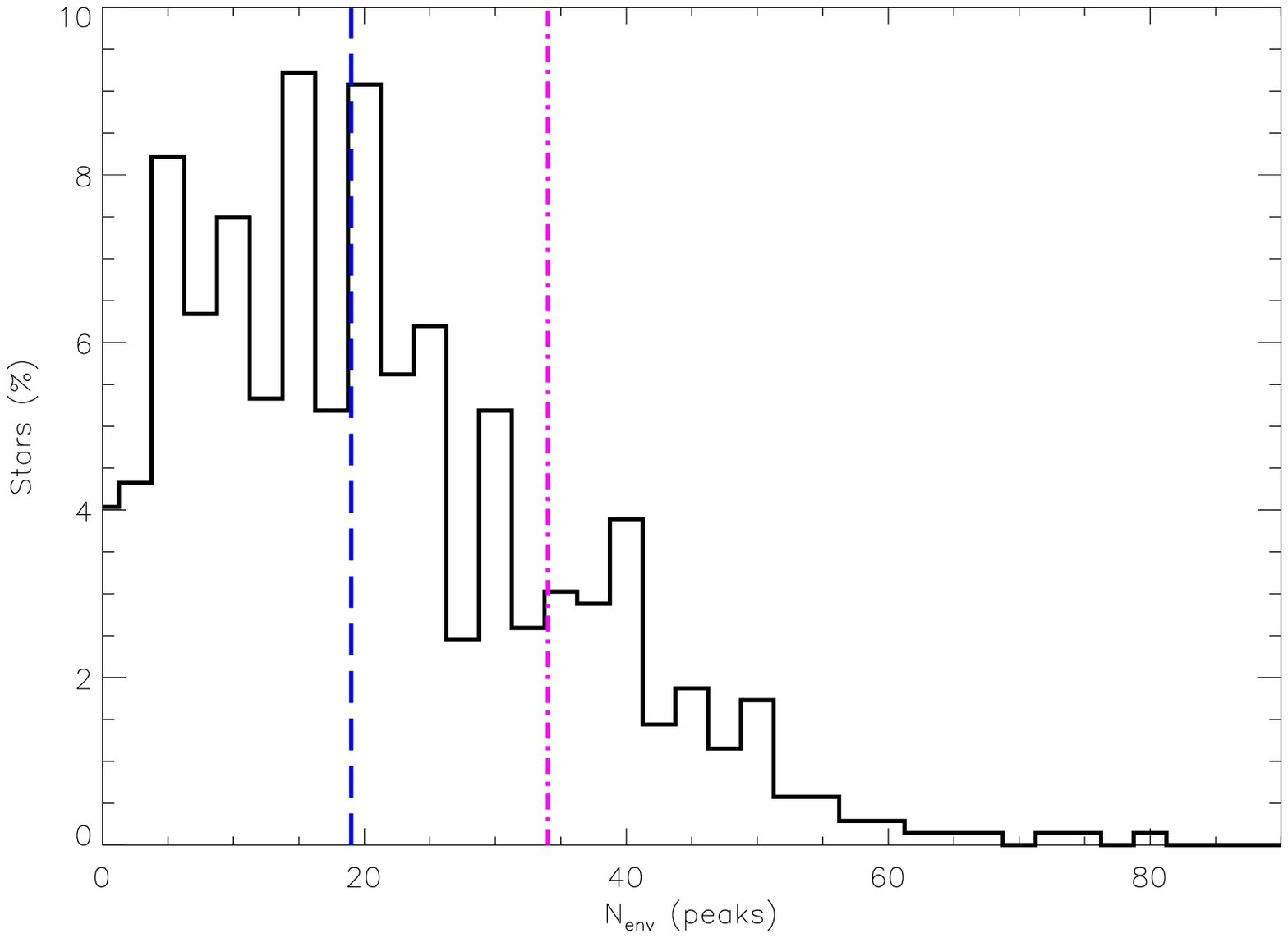}
\includegraphics[width=0.495\textwidth]{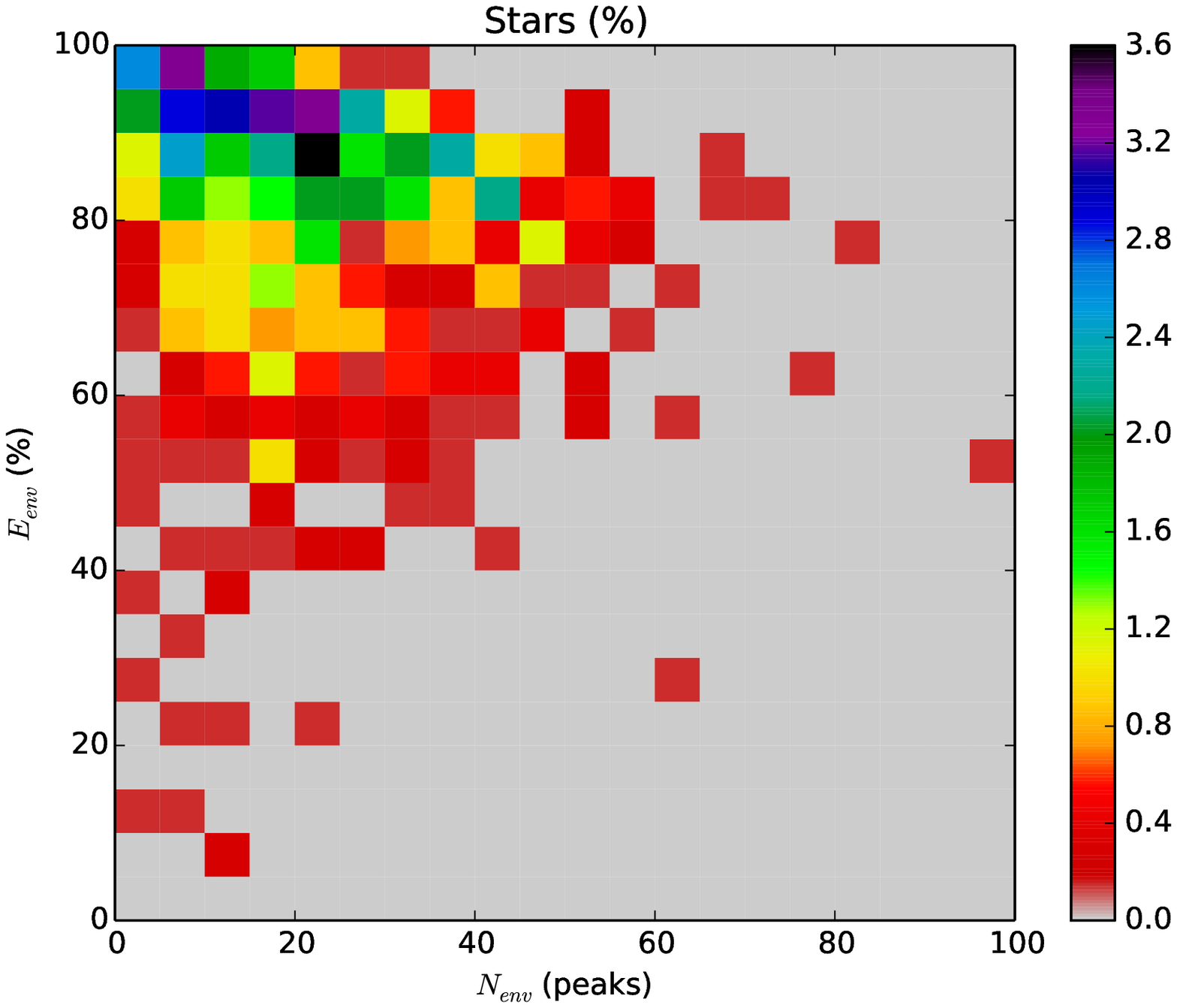}
\caption{Left panel: Distribution of stars according to the number of peaks in their envelope. The mean number of peaks of the envelope (blue dashed line) is similar to the estimate made by \citet{Lignieres2009} for a \dScu star model rotating at a rate of 0.59 $\Omega_{K}$ (purple dashed-dotted line). Right panel: Distribution of stars according to the number of peaks and the energy of their envelope.}
\label{f:SNenv}
\end{figure*}

\begin{figure*}
\centering
\includegraphics[width=0.495\textwidth]{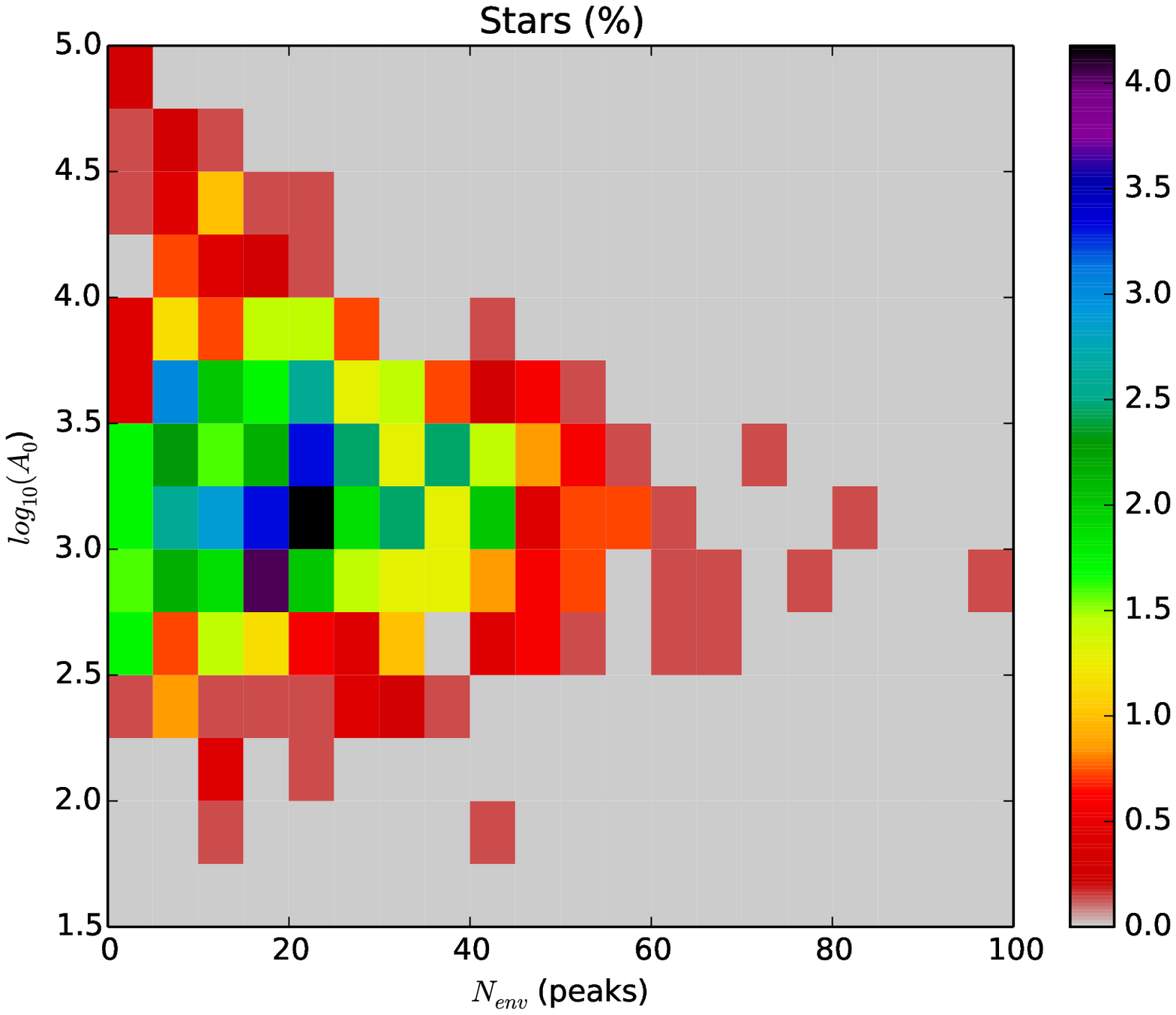}
\includegraphics[width=0.495\textwidth]{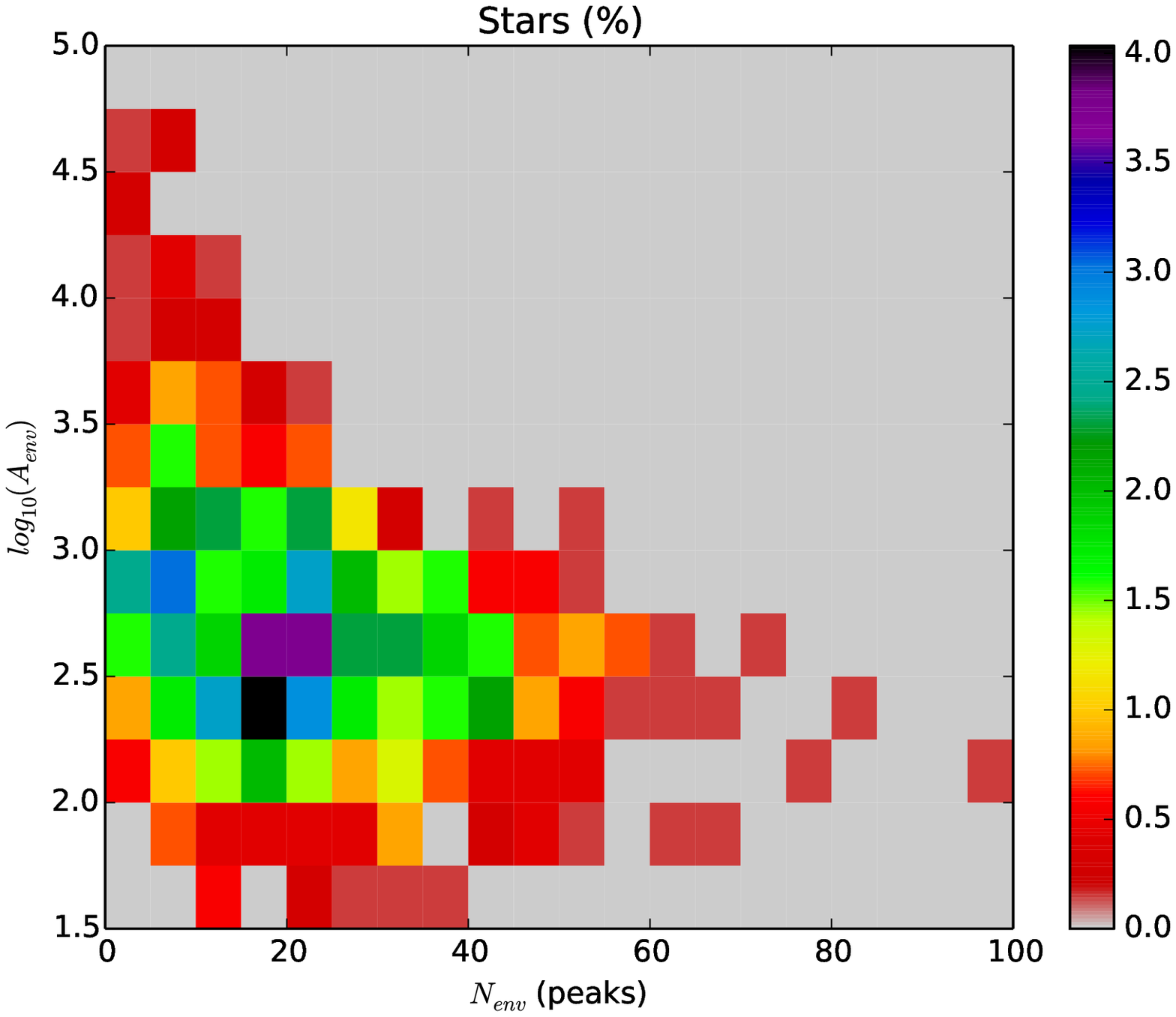}
\caption{Left panel: Distribution of stars according to the number of peaks and the amplitude of their highest-amplitude mode. Right panel: Distribution of stars according to the number of peaks and the mean amplitude of their envelope.}
\label{f:amp}
\end{figure*}

\begin{figure*}
\centering
\includegraphics[width=0.495\textwidth]{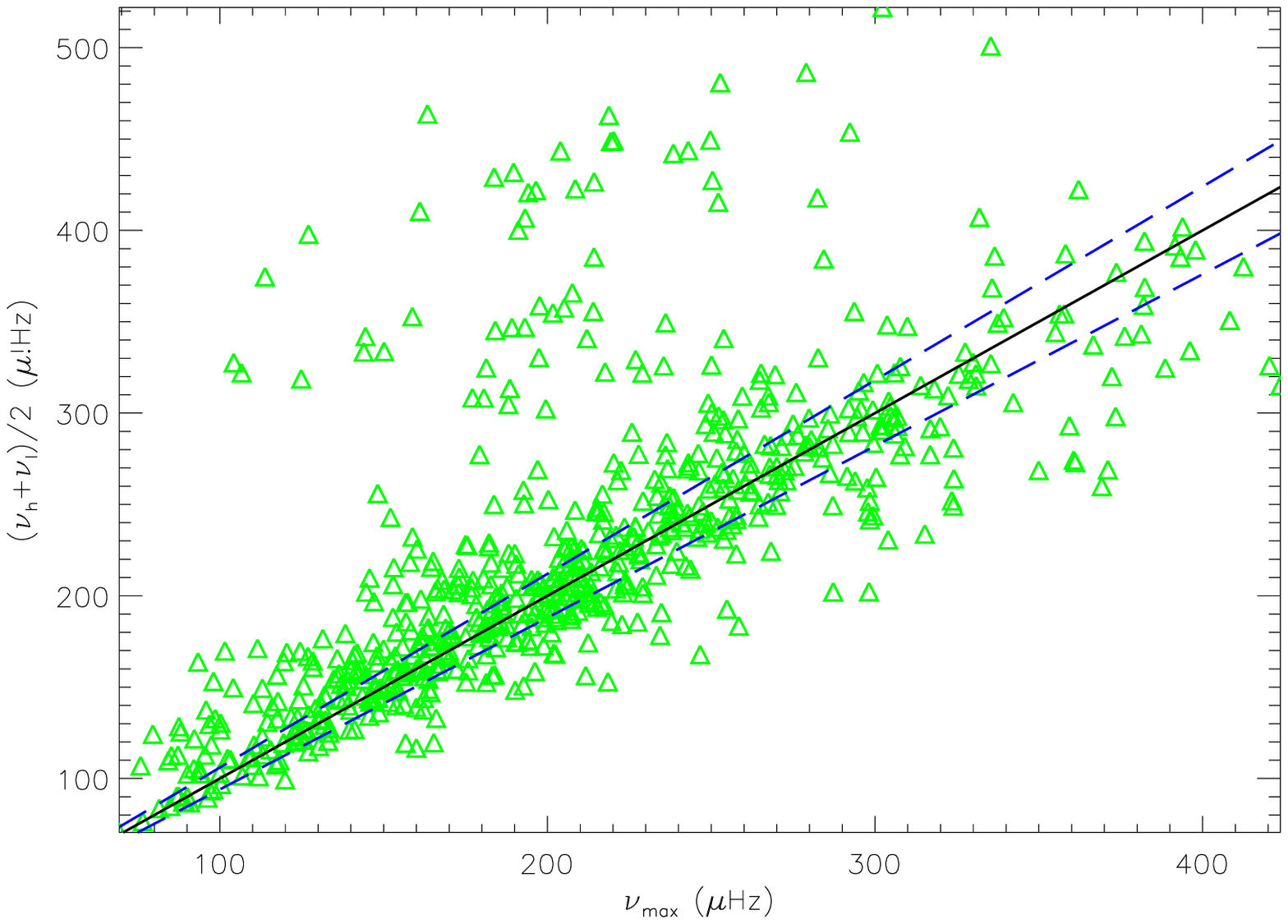}
\includegraphics[width=0.495\textwidth]{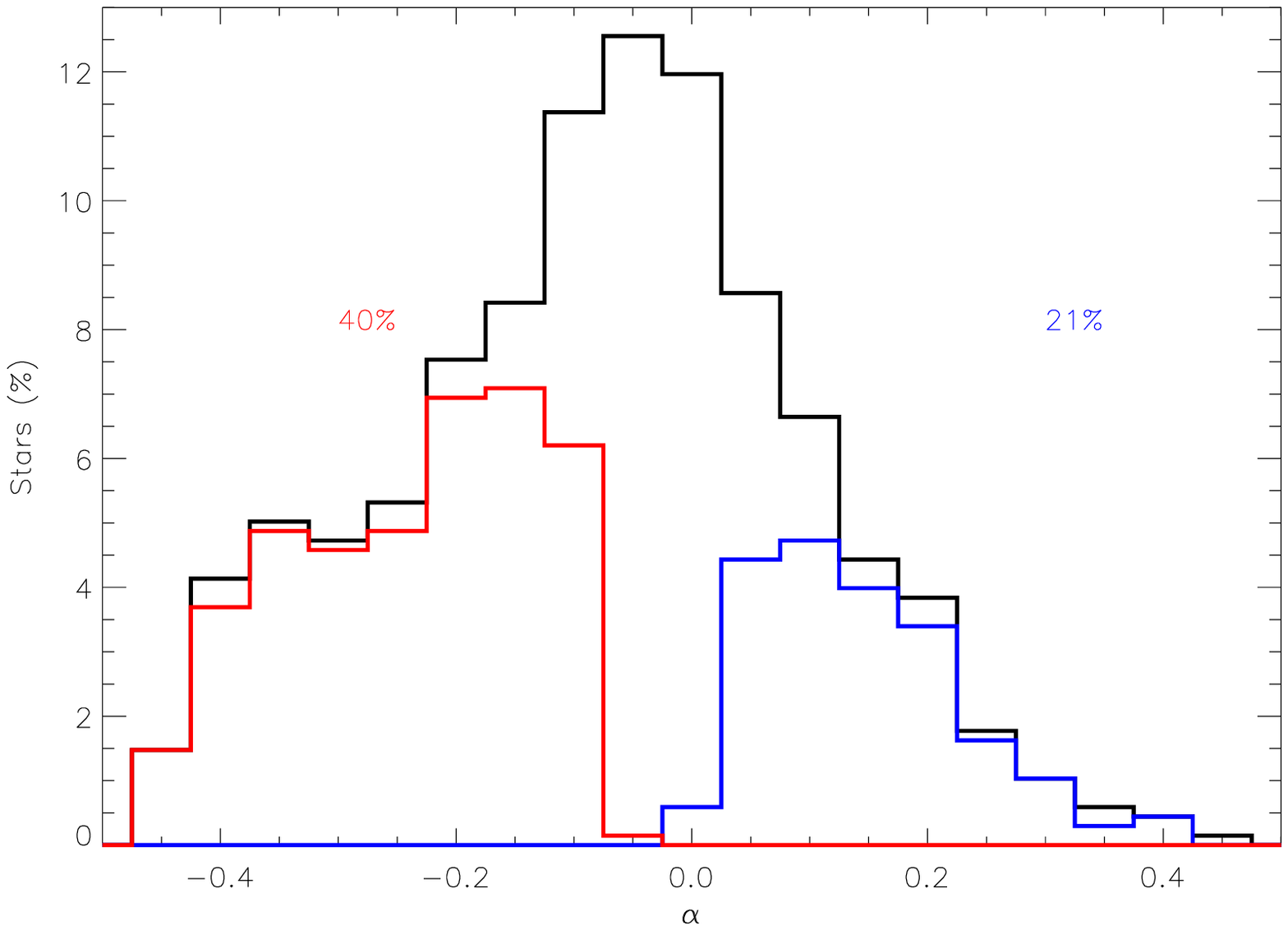}
\caption{Left panel: Relation between $\nu_{\rm max}$ and the mean frequency of the envelope for \dScu stars (green triangles). The black line points to the 1:1 relation. Blue dashed lines are the limits for a significant departure from the solar case (3$\sigma$, see text). Right panel: Distribution of stars according to their asymmetry (black histogram). Blue (Red) histogram denotes the proportion of stars whose $\nu_{\rm max}$ deviation from the mean frequency of the envelope towards lower (higher) frequencies is significantly higher than the solar case (see text).}
\label{f:Asym}
\end{figure*}

\subsection{Number of peaks}
\label{ss:nop}

As we can see in Fig.~\ref{f:SNenv} (left), the power spectra envelopes of these \dScu stars are formed by a few tens of peaks. The observed typical values of the number of modes are between 5 to 45 modes and the mean value is 20 modes. The estimated number of 2-period island modes for a \dScu star model rotating at 0.59 $\Omega_{K}$ \citep[see][]{Lignieres2009} is higher ($34 \pm 2$ modes) but it is inside the typical regime. Those values are also in agreement with the number of modes found for CID~546\footnotemark[1]\footnotetext[1]{We abbreviate the nomenclature of CoRoT~ID to CID not taking into account the zeroes at the left of its identification number of ten digits.}, CID~8669, and KIC~5892969 \dScu stars \citep{BarceloForteza2017}.\\

The proportion of stars with a lower number of high energy peaks might be explained in terms of a selection mechanism, such as trapping or resonances \citep[e.g.][]{Dziembowski1982,Dziembowski1990}. In addition, 2-period island modes show lower amplitude peaks with higher inclinations. Therefore, those 2-period island modes have amplitudes comparable to the chaotic modes for stars with inclinations close to the equator \citep{Lignieres2009}. As a consequence, the lower 2-period island modes could not have enough energy, $E_{i}$, to be considered as located inside the envelope. The energy of the envelope, $E_{\rm env}$ represents an 80\% or higher proportion of the entire signal for most \dScu stars. We observed a decrease in the number of stars with high number of peaks towards lower $E_{\rm env}$ (see Fig.~\ref{f:SNenv}, right). This is in agreement with the scenario described before.\\

We can also observe a large tail to higher number of peaks. This phenomenon can be explained by the long duration of the \textit{Kepler} campaigns, which allowed us to observe the long-period cyclic variations and their associated high-amplitude sidelobes \citep[e.g.,][]{BarceloForteza2015}. In most cases ($\sim 70 \%$), those sidelobes are split from the main peak only a few tens of nHz or less and their characteristics are those predicted for resonance \citep[e.g.,][]{Moskalik1985}, binarity \citep{ShK2012} or aliasing \citep{Murphy2013}. The results presented in \cite{Bowman2016} are in agreement with our observations. Then, we did not take the sidelobes of the modes into account to calculate the typical number of peaks of the envelope.\\

\subsection{Asymmetry}
\label{ss:asym}

The typical amplitudes of the highest modes ($A_{0}$) of \dScu stars are around thousands of ppm (see Fig.~\ref{f:amp}, left). However, the mean amplitude of the envelope is almost one order of magnitude lower $A_{\rm env} \sim 300$ ppm (see Fig.~\ref{f:amp}, right). We also observed some cases with amplitudes up to $10^5$ ppm and low number of modes that have been classified as HADS (see Sect.~\ref{ss:hads}).\\

An interesting characteristic of the envelope is its asymmetry, which has been calculated for red giants by \citet{Kallinger2010a}. They find a weak asymmetry around 3.1$\pm$1.3\% of the $\nu_{max}$ value; such a difference between the weighted frequency and the mean frequency of the envelope is similar to that found in the solar case. We have calculated the asymmetry of the envelope for \dScu stars, $\alpha$ (see Eq.~\ref{e:asym}), and we observed that their envelope can be highly asymmetric, up to $|\alpha|$=0.45. Specifically, 61\% of stars have a significantly higher asymmetry than the solar case (see Fig.~\ref{f:Asym}) considering a higher difference than $3\sigma$.\\

This asymmetry is not produced by a bias due to aliasing. The Nyquist frequency of \textit{CoRoT} data is higher than the typical frequencies of \dScu type stars, and a superNyquist analysis \citep{Murphy2013} has been done with \textit{Kepler} data for the peaks of the envelope. There are several possible causes such as a selection mechanism \citep[resonance, trapping; e.g.][]{Dziembowski1982,Dziembowski1990}, or the variation of the large separation with frequency as is also seen in other kinds of stars \citep[e.g.][]{Mosser2010}.\\

The number of stars showing asymmetry towards low frequency modes is much higher (41\%) than those with high frequency asymmetry in their envelope (21\%, see Fig.~\ref{f:Asym}, right). This difference could be a consequence of the lower energy required to excite these lower frequency modes compared with the energy required by the higher frequency ones. Therefore, the asymmetry in the envelope may be produced by an asymmetry in the growth rate of the modes.\\

These statistical properties are not significantly modified by the definition of $\nu_{\rm max}$. To calculate the asymmetry we can also use the frequency of the highest amplitude mode ($\nu_{0}$) or the weighted mean frequency of the power spectrum instead of the amplitudes (see Eq.~\ref{e:nup})
\begin{equation}
 \nu_{\rm max}' = \frac{\sum^{N_{\rm env}}_{i=0} A_{i}^2 \nu_{i}}{\sum^{N_{\rm env}}_{i=0} A_{i}^2} \, .
\label{e:nup2}
\end{equation}
For both cases, we observed the same number of stars with asymmetric envelopes. Moreover, there are the same proportion of stars with asymmetry towards low and high frequencies although the extreme cases have higher $|\alpha|$. This is in agreement with our hypothesis of the energy requirements to excite the modes.

\subsection{Frequency at maximum power}
\label{ss:umax}

\begin{figure}
\centering
\includegraphics[width=.5\textwidth]{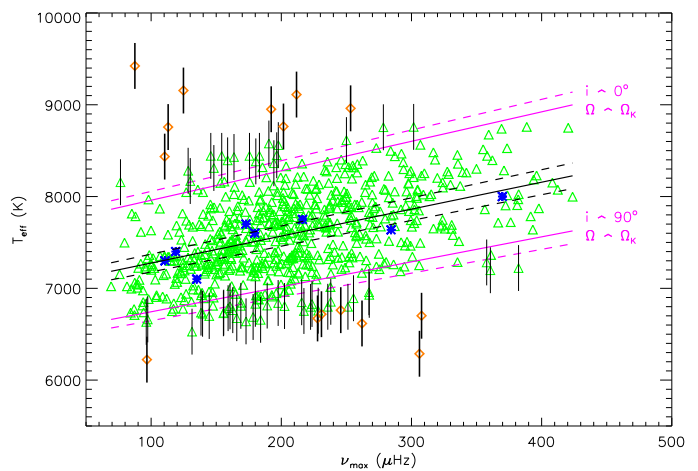}
\caption{Relation between $\nu_{\rm max}$ and $\bar{T}_{\rm eff}$ for \textit{CoRoT} Sismo-channel and \textit{Kepler} \dScu stars (blue asterisks, and green triangles, respectively). The black line represents the fitted relation. Purple lines mark the limits of the predicted dispersion due to the gravity-darkening effect. All dashed lines represent the estimated error of the linear fit. Orange diamonds correspond to the 2\% of \textit{Kepler} \dScu stars outside the limits when taking into account only 1$\sigma$ error of their $T_{\rm eff}$ measurement (see text). We show only a few error bars for clarity.}
\label{f:Turel}
\end{figure}

\begin{table}
\caption{Parameters of the fitted scaling relation taking into account different definitions of the frequency at maximum power.}
\label{t:tnu}
\centering
\begin{tabular}{c c c c c c}
                            & Slope & Y-intercept& $\sigma$& R & $P_{u}$\\
Def.\tablefootmark{a} & (K/$\mu$Hz) &$\pm 50$ (K)&(\%)&& ($10^{-30}$\%)\\
\hline
$\nu_{\rm max}$ &  2.94 $\pm$ 0.24&  6980 & 5.82&  0.424& 7\\
$\nu_{0}$       &  2.39 $\pm$ 0.20&  7110 & 5.87&  0.408& 1680\\
$\nu_{\rm max}'$&  2.76 $\pm$ 0.23&  7024 & 5.84&  0.421& 23\\
\hline 
\end{tabular}
\tablefoot{\tablefoottext{a}{We tested the following definitions: the weighted mean frequency ($\nu_{\rm max}$, see Eq.~\ref{e:nup}), the frequency of the highest amplitude mode ($\nu_{0}$), and the weighted mean frequency of the power spectrum ($\nu_{\rm max}'$, see Eq.~\ref{e:nup2}).}}
\end{table}

Having determined the values of $\nu_{\rm max}$, we can get the effective temperatures $T_{\rm eff}$ from the input catalogues of both satellites and make a plot as shown in Fig.~\ref{f:Turel}. Using only the values of $T_{\rm eff}$ from the \textit{Kepler} catalogue \citep{Brown2011,Huber2014}, we obtained the following linear relation (see Fig.~\ref{f:Turel}):
\begin{equation}
\bar{T}_{\rm eff} (K) = (2.94 \pm 0.24) \cdot \nu_{\rm max} (\mu Hz) + (6980 \pm 50) \, .
\label{e:Tnumax}
\end{equation}
The observed dispersion of the data is around $\sigma \sim 6$\% and the value of the Pearson correlation coefficient is $R=0.42$, meaning a probability of being uncorrelated \citep[$P_{u}$; i.e.][]{Taylor1997} around
\begin{equation}
P_{u} = \frac{2 \Gamma \left( \frac{N-1}{2} \right)}{\sqrt{\pi}\Gamma \left(\frac{N-2}{2} \right)} \int^{1}_{R} \left(1-r^2\right)^{\frac{N-4}{2}} dr \approx 7.1 \times 10^{-30} \% \, ,
\label{e:Pu}
\end{equation}
where $\Gamma \left( x \right)$ is the Gamma function, and $N$ is the number of \dScu stars of the sample. This is the probability that $N$ measurements of a priori two uncorrelated variables gives $|R| \geq  0.42$. Taking into account that a probability $P_{u} \leq 5 $\% and $P_{u } \leq 1 $\% mean a significant and a highly significant correlation, respectively, we found a highly significant evidence of a linear relation between the frequency at maximum power and the effective temperature. Moreover, notice that this relation is also followed by the stars observed with the \textit{CoRoT} sismo-channel (see Fig.~\ref{f:Turel} blue asterisks) that have been studied in detail by several authors \citep[e.g.][]{BarceloForteza2017}.\\

We tested other possible scaling relations such as $\bar{T}_{\rm eff}-\nu_{0}$ or $\bar{T}_{\rm eff}-\nu_{\rm max}'$ (see Eq.~\ref{e:nup2}). Although all them have similar slope and y-intercept values, the relation we found in Eq.~\ref{e:Tnumax} has the highest Pearson coefficient and lower $P_{u}$ (see Table~\ref{t:tnu}).

\begin{figure}
\centering
\includegraphics[width=.5\textwidth]{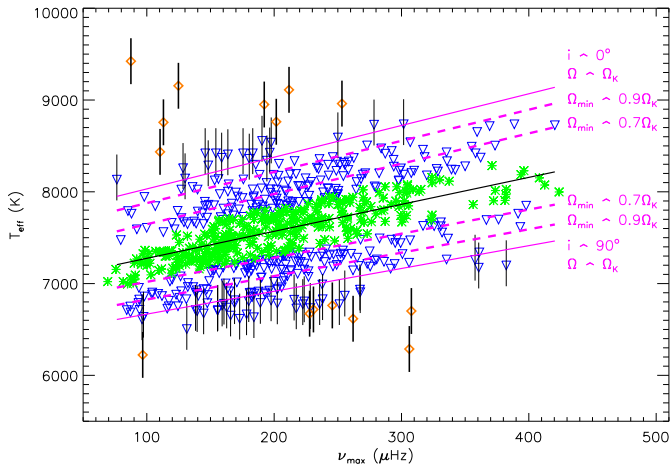}
\caption{$\nu_{\rm max}$-$\bar{T}_{\rm eff}$ relation (black line) with the different dashed purple lines representing the lower limit of the rotation ratio $\Omega/\Omega_{K}$. Solid purple lines mark the limits of the predicted dispersion due to the gravity-darkening effect. Blue inverted triangles represent the extreme fast \dScu stars observed by \textit{Kepler}. Green asterisks represent those stars with inclination-rotation degeneracy (see text). Orange diamonds correspond to the 2\% of \textit{Kepler} \dScu stars outside the limits when taking into account only 1$\sigma$ error of their $T_{\rm eff}$ measurement (see text). We show only a few error bars for clarity reasons.}
\label{f:Omega_min}
\end{figure}

\section{Discussion}
\label{s:gdeex}

The $\bar{T}_{\rm eff}-\nu_{\rm max}$ scaling relation might be explained by the excitation of higher radial order peaks with higher temperature, as was predicted by \citet[][see Fig.~2 in their publication]{Dziembowski1997}. \citet{Balona2011} look for a dependence between the frequency of highest amplitude and the effective temperature. Their predicted frequency of highest growth rate seem to increase with the effective temperature (see Fig.~2 in their publication). They find that the measured values vary widely respect the predicted ones, however, the maximum value of the measured temperature increases with the frequency. We observed the same effect when hybrid stars are included in our analysis like \cite{Balona2011} do in their study.\\

There are several possible causes to explain the observed dispersion in the scaling relation such as the error of the temperature measurements, a possible systematic error derived of the definition of the frequency at maximum power, or a physical mechanism such as the gravity-darkening effect. The relative error of the measured effective temperatures of the Kepler catalogue are around the 3\% and it is not enough to explain the observations ($\sigma \sim 6$\%, see Section~\ref{ss:umax}). Moreover, there is non significant variation of the observed dispersion when we use $\nu_{0}$ or $\nu_{\rm max}'$ instead of $\nu_{\rm max}$ ($\Delta \sigma\sim 0.05 $\% and $\sim 0.01$\%; see Table~\ref{t:tnu}).\\

\begin{figure}
\centering
\includegraphics[width=.5\textwidth]{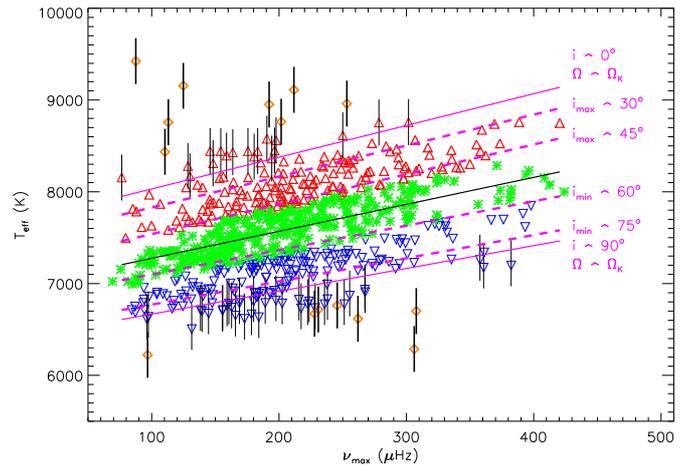}
\caption{Same as Fig.~\ref{f:Omega_min} but the different dashed purple lines represent the higher (lower) limits of the inclination $i$. Red (blue inverted) triangles represent the extreme fast \dScu stars observed by \textit{Kepler}.}
\label{f:i_lims}
\end{figure}

We might explain the dispersion with the gravity-darkening effect \citep{vonZeipel1924}:
\begin{equation}
T_{\rm eff}^{4}(i) = C \cdot g_{\rm eff}^{\beta}(i) \; ,
\label{e:g-deff}
\end{equation}
where $T_{\rm eff}(i)$ and $g_{\rm eff}(i)$ are the effective temperature and the effective gravity observed for a given inclination $i$, $C$ is a constant, and the value of $\beta$ depends of the evolutionary stage of the star \citep{Claret1998}. Assuming that the surface of the star is a Roche surface \citep{PerezHernandez1999}, the value of the effective gravity is 
\begin{equation}
g_{\rm eff} \approx g- R(i) \Omega^{2} \sin^{2} \{ i \} \; ,
\label{e:geff}
\end{equation}
where, $R(i)$ is the radius of the star for an inclination $i$, and $\Omega$ is its rotation. Therefore, the gravity-darkening effect between the equator and the poles can be obtained,
\begin{equation}
\delta T_{\rm eff} = \frac{T_{\rm eff,p}-T_{\rm eff,e}}{T_{\rm eff,p}} \approx 1-\left( 1- \frac{5}{3}\epsilon^{2}+ \frac{5}{6}\epsilon^{4} \right)^{\frac{\beta}{4}} \; ,
\label{e:dteff}
\end{equation}
where the value of $T_{\rm eff,p}$ and $T_{\rm eff,e}$ are the effective temperatures at the pole and at the equator of the stars, respectively, and \begin{equation}
\epsilon^{2} = \frac{\Omega^{2} R^{3}}{G M} \; ,
\label{e:eps2}
\end{equation}
where $M$ is the mass, and $R$ is the radius of a star with spherical symmetry. Moreover, the temperature difference between mid-latitudes, $i \sim 55^{\circ}$, and any other inclination value is
\begin{equation}
\delta \bar{T}_{\rm eff}(i) = \frac{T_{\rm eff}(i)-\bar{T}_{\rm eff}}{\bar{T}_{\rm eff}}  \approx \left( \frac{1-\left( \frac{R(i)}{R} \right)^{3} \epsilon^{2} \sin^2 \{ i \} }{1- (2/3)\cdot \epsilon^{2}} \right)^{\frac{\beta}{4}} -1 \; .
\label{e:dteffi}
\end{equation}
In fact, $T_{\rm eff}(i \sim 55^{\circ}) \approx \bar{T}_{\rm eff}$ since at that inclination the effective surface gravity is the same as that of the spherically symmetric star \citep{PerezHernandez1999}.\\

\subsection{Extreme fast rotators}
\label{ss:fast}

The \dScu stars with the highest departure from the $\bar{T}_{\rm eff}$-$\nu_{\rm max}$ relation should be those equator-on or pole-on stars whose rotation rate is close to their break-up frequency,
\begin{equation}
\Omega \approx \Omega_{K}=\sqrt{ \frac{8 G M}{27 R_{\rm p}^{3}}} \, ,
\label{e:OmegaK}
\end{equation} 
where $R_{p}$ is the polar radius
\begin{equation}
R_{p}=\frac{R}{1+\epsilon^{2}/3} \, .
\label{e:Rp}
\end{equation} 
Then, Eq.~\ref{e:eps2} is equivalent to
\begin{equation}
\epsilon^{2} = \left( \frac{\Omega}{\Omega_{K}}\right)^{2} \epsilon_{\rm max}^{2}\; ,
\label{e:eps2bis}
\end{equation}
where the maximum value possible for this dimensionless parameter is $\epsilon_{\rm max}^{2} \sim 0.451$. Taking into account that $\beta \sim 1$ for stars with fully radiative envelope \citep{Claret1998}, the maximum temperature variation due to the gravity-darkening effect is
\begin{equation}
\delta \bar{T}_{\rm eff}(0^{\circ}) \approx 9.4\% \; ; \; \delta \bar{T}_{\rm eff}(90^{\circ}) \approx -7.3\% \; .
\label{e:gdisp}
\end{equation} 
Therefore, the predicted dispersion includes 98\% of the stars, which suggests that the gravity-darkening effect might explain the real dispersion of the observations (see Fig.~\ref{f:Turel}).\\ 

Using only the observed temperature departure,
\begin{equation}
dT_{\rm eff,obs} = \frac{T_{\rm eff}-\bar{T}_{\rm eff}}{\bar{T}_{\rm eff}} \; ,
\label{e:dteffobs}
\end{equation} 
and Eq.~\ref{e:dteffi}, we can estimate the minimum value of the rotation rate ($\Omega_{\rm min}$) for each star, assuming it is pole-on or equator-on and fulfils the condition 
\begin{equation}
|dT_{\rm eff,obs}| > ET_{\rm eff}/\bar{T}_{\rm eff} \approx 3\% \; ,
\label{e:econd}
\end{equation} 
where $ET_{\rm eff}$ is the error of the measured effective temperature (see Fig.~\ref{f:Omega_min}). These stars (blue inverted triangles) should then be extreme fast rotators because a high rotation rate is needed to fulfil this last condition ($\Omega \geq 0.7 \Omega_{K}$). In contrast, we can also delimit the inclination of these \dScu stars by assuming $\Omega \sim \Omega_{K}$ and using $dT_{eff,obs}$ in Eq.~\ref{e:dteffi}. We could find those limits for extreme rotators with inclinations $i\lesssim 45^{\circ}$ or $i\gtrsim 60^{\circ}$ (see red triangles, and blue inverted triangles in Fig.~\ref{f:i_lims}, respectively). There is degeneracy between $i$ and $\Omega$ for those stars that do not satisfy Eq.~\ref{e:econd}, (green astersiks). Nevertheless, a deeper study of each power spectrum should allow us to differentiate between a moderate or slow rotator ($\forall i$ and $\Omega < 0.7 \Omega_{K}$), from a extreme rotator with an inclination close to the mid-latitude \citep[$i\sim 55^{\circ}$ and $\forall \Omega$; see ][]{BarceloForteza2017}. The limits of rotation and inclination for all \dScu stars can also be found in the online material (Table~A.1).\\

\begin{table*}
\caption{Measured (8th column) and calculated (7th column) effective temperature for well-known \dScu stars (see text).}
\label{t:teff}
\centering
\begin{tabular}{c c c c c c c c c}
N & CID & $\frac{\Omega}{\Omega_{K}}$ (\%)\tablefootmark{a} & $i$ $( ^{\circ})$\tablefootmark{a}& $\nu_{\rm max}$ ($\mu$Hz)\tablefootmark{b}& $\bar{T}_{\rm eff}$ (K)\tablefootmark{b}& $T_{\rm eff}(i)$ (K)\tablefootmark{b}& $T_{\rm eff}$ (K)\tablefootmark{a} & References\\ 
\hline

1 & 1043 &  25 $\pm$ 3   & 70 $\pm$ 5       & 111 $\pm$ 4  & 7280 $\pm$ 180 & 7270 $\pm$ 180 & 7300 $\pm$ 200 & \cite{Hareter2014}         \\
2 & 123  &  38 $\pm$ 2   & 76 - 82          & 119 $\pm$ 8\tablefootmark{c}  & 7320 $\pm$ 180 & 7310 $\pm$ 130 & 7510 $\pm$ 130 & \cite{Chen2016}         \\
3 & 8669 &  70 $\pm$ 7   & 55 - 90          & 135 $\pm$ 6  & 7400 $\pm$ 190 & 7250 $\pm$ 210 & 7000 $\pm$ 200 & \cite{BarceloForteza2017}  \\
4 & 5685 &  26           & 45               & 173 $\pm$ 4  & 7600 $\pm$ 220 & 7600 $\pm$ 230 & 7700 $\pm$ 200\tablefootmark{d} & \cite{Escorza2016} \\
5 & 546  &  31 $\pm$ 3   & 15 - 35          & 179 $\pm$ 11 & 7630 $\pm$ 230 & 7660 $\pm$ 250 & 7660 $\pm$ 250 & \cite{BarceloForteza2017}  \\
6 & 8170 &  50 $\pm$ 5   & 73.2 $\pm$ 0.6   & 216 $\pm$ 5\tablefootmark{c} & 7830 $\pm$ 250 & 7780 $\pm$ 260 & 7750 $\pm$ 250 & \cite{Creevey2009} \\
7 & 7528 &  60 $\pm$ 10  & 45 - 70          & 284 $\pm$ 9  & 8200 $\pm$ 300 & 8100 $\pm$ 350 & 7600 $\pm$ 200 & \cite{GarciaHernandez2013} \\
8 & 7613 & 100 $\pm$ 14\tablefootmark{e}  & 51 $\pm$ 2       & 370 $\pm$ 12 & 8300 $\pm$ 230 & 8400 $\pm$ 280 & 8000 $\pm$ 200 & \cite{GarciaHernandez2009} \\ \hline 
\end{tabular}
\tablefoot{\tablefoottext{a}{Those parameters have been taken from the literature. The authors are listed in the "Reference" column.} \tablefoottext{b}{The frequency at maximum power ($\nu_{\rm max}$), the mean effective temperature ($\bar{T}_{\rm eff}$), and the effective temperature for a given inclination ($T_{\rm eff}(i)$) have been calculated using Eqs.~\ref{e:nup},~\ref{e:Tnumax}, and~\ref{e:g-deff}, respectively (see text).} \tablefoottext{c}{The frequencies and the amplitudes of the modes to calculate $\nu_{\rm max}$ of CID~123 and CID~8170 have been taken from \cite{Poretti2009} and \cite{Costa2007}, respectively.} \tablefoottext{d}{The effective temperature of CID~5685 have been taken from \cite{Poretti2005}.} \tablefoottext{e}{We obtained the rotation rate of CID~7613 studying the power spectrum of the star (see text).}}
\end{table*}

\subsection{The case of the well-known \dScu stars}
\label{ss:wellkn}

We tested the $\bar{T}_{\rm eff}$-$\nu_{\rm max}$ relation and the gravity-darkening effect using the known data of eight previously studied stars (see Table~\ref{t:teff}). We have chosen those \dScu stars because their frequency at maximum power range spans all the studied region. This region includes $\nu_{\rm max}$ higher than the Nyquist frequency of the LC Kepler light curves in order to test our results in this regime too. In addition, this sample has stars with a wide range of inclinations ($i\sim 25-84^{\circ}$) and rotation rates ($\Omega/\Omega_{K}\sim 6-100\%$). We calculated their rotation rate using Eq.~\ref{e:OmegaK} and the values of mass, radius, rotation, inclination and/or projected velocity ($v \sin i$) provided by the different authors (see Table~\ref{t:teff}).\\ 

\begin{figure}
\centering
\includegraphics[width=0.5\textwidth]{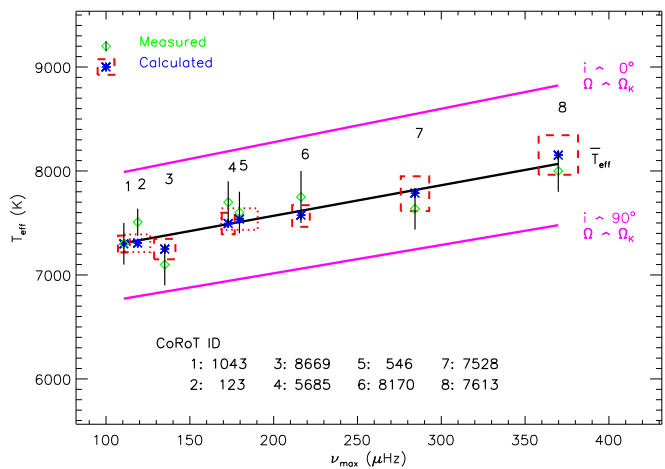}
\caption{Measured (green diamonds) and calculated (blue asterisks) effective temperature for \textit{CoRoT} Sismo-channel \dScu stars. Red and purple lines denote the $\bar{T}_{\rm eff}-\nu_{\rm max}$ relation and the limits of the dispersion due to gravity-darkening effect (see text).}
\label{f:WKDS}
\end{figure}

Two of these \dScu stars, CID~1043 (star 1) and CID~8170 (star 6), are well-known spectroscopic binaries whose structural and orbital parameters were studied by \cite{Hareter2014} and \cite{Creevey2009}, respectively. The \dScu star with chemical peculiarities studied by \cite{Escorza2016}, CID~5685 (star 4), is a good example of the observed difference of temperature when two different kinds of methods are used. The authors use spectroscopic observations to obtain $T_{\rm eff}=7200 \pm 80$ K, quite different from that obtained with Str\"omgren $uvby\beta$ photometry \citep[$T_{eff}=7700 \pm 200$ K;][]{Poretti2005}. This difference is consistent with the observations of \cite{Pinsonneault2012} who, in their study, found differences of the order of 250 K between the spectroscopic and photometric methods. Another method to obtain the effective temperature is the frequency modelling. In this way, \cite{Chen2016} studied CID~123 (star 2) finding an effective temperature around $T_{\rm eff} = 7510 \pm 130$ K for Z=0.008 and $T_{\rm eff} \sim 7380$ K for Z=0.009. In the worst scenario, this result still being inside the 1-$\sigma$ error for both \citet[][$T_{\rm eff} = 7400 \pm 200$ K]{Poretti2009} and our results (see Table~\ref{t:teff}). In any case, it is preferable to compare with those temperatures obtained with the same method used to calibrate the $\bar{T}_{\rm eff} - \nu_{\rm max}$ relation. In the case of CID~7613 (star 8), \cite{GarciaHernandez2009} obtained a minimum rotation around 20 $\mu$Hz. To obtain $\Omega/\Omega_{K}$, we looked for the highest peak of the power spectrum of the subspectrum (see Fig.~3 of their publication) inside the regime
\begin{equation}
\frac{v \sin i}{R} \leq \Omega \leq \Omega_{K} \; ,
\label{e:slim}
\end{equation}
i. e., $\frac{\Omega}{2 \pi} \in \left[20,25.8\right]$ $\mu$Hz. To calculate its break-up frequency, we used Eq.~\ref{e:lsep_agh} with $\Delta \nu = 53$ $\mu$Hz. Therefore, this star has a rotation rate equal to its break-up frequency and equal to half its large separation. This is in agreement with the $\Omega \sim 0.94 \Omega_{K}$ model of a \dScu star calculated by \cite{Reese2017}\footnotemark[2]\footnotetext[2]{The authors use $\Omega_{C}$ instead of $\Omega_{K}$ (see Eq.~\ref{e:OmegaC}, and~\ref{e:OmegaK}, respectively) as a reference to calculate their models. Therefore, a model with $\Omega \sim 0.94 \Omega_{K}$ is equivalent to $\Omega \sim 0.7 \Omega_{C}$}.\\

Once we know all those structural parameters ($\Omega / \Omega_{K}$, $i$, and $T_{\rm eff}$), we obtain the value of the calculated effective temperature ($T_{\rm eff}(i)$, see Table~\ref{t:teff}) using Eq.~\ref{e:Tnumax}, and~\ref{e:dteffi}. To calculate $\bar{T}_{\rm eff}$, we only need the value of $\nu_{\rm max}$, i.e., the frequencies and amplitudes of the pulsations. These parameters of the modes are also provided by the different authors.\\

Finally, the value of the calculated effective temperature matches the measured effective temperature inside the 1-$\sigma$ error for all stars studied (see Fig.~\ref{f:WKDS} and Table~\ref{t:teff}). On the one hand, those low- or moderate-rotation rate \dScu stars of the \textit{CoRoT} sample (see stars 1, 2, 4, 5, and 6) do not have a significant difference of temperature owing to the inclination. Therefore, those stars allow us to suggest the validity of the $\bar{T}_{\rm eff}-\nu_{\rm max}$ relation. On the other hand, high-rotation rate \dScu stars (3, 7, and 8) present a higher dependence with the inclination. In those three cases, the higher limit of inclination would be more appropriate to explain the observed temperature.\\

\begin{table}
\caption{Rotation rate limits of the outsiders in our sample of stars.}
\label{t:out}
\centering
\begin{tabular}{c c c c c c}
KIC & $\nu_{\rm max}$ ($\mu$Hz)& $\bar{T}_{\rm eff}$ (K)& $T_{\rm eff}$ (K)\tablefootmark{a} & $\frac{\Omega}{\Omega_{K}}$ (\%)\\ 
\hline
 1577912 &  250 $\pm$    10& 7700 $\pm$   100& 8960&$\gtrsim$ 112 \\
 2554867 &  113 $\pm$     1& 7310 $\pm$    80& 8760&$\gtrsim$ 122 \\
 3347643 &  310 $\pm$    20& 7900 $\pm$   100& 6290&$\gtrsim$ 113 \\
 5904699 &  246 $\pm$     8& 7700 $\pm$   100& 6760&$\gtrsim$ 102 \\
 7051690 &  262 $\pm$     9& 7800 $\pm$   100& 6620&$\gtrsim$ 106 \\
 7457355 &  231 $\pm$     5& 7700 $\pm$   100& 6720&$\gtrsim$ 102 \\
 7538950 &  200 $\pm$    20& 7600 $\pm$   100& 8760&$\gtrsim$ 111 \\
 7765684 &  146 $\pm$     9& 7410 $\pm$    90& 8440&$\gtrsim$ 116 \\
 8500899 &  228 $\pm$     9& 7700 $\pm$   100& 6670&$\gtrsim$ 103 \\
 8579615 &  110 $\pm$     6& 7310 $\pm$    80& 8440&$\gtrsim$ 110 \\
 9115188 &  308 $\pm$     7& 7900 $\pm$   100& 6700&$\gtrsim$ 107 \\
10273384 &   97 $\pm$     6& 7270 $\pm$    80& 6220&$\gtrsim$ 106 \\
10788451 &  190 $\pm$    10& 7550 $\pm$   100& 8950&$\gtrsim$ 120 \\
11027806 &  120 $\pm$    70& 7350 $\pm$    80& 9160&$\gtrsim$ 133 \\
11081490 &   88 $\pm$     7& 7240 $\pm$    70& 9420&$\gtrsim$ 142 \\
11823661 &  210 $\pm$    30& 7600 $\pm$   100& 9110&$\gtrsim$ 123 \\
\hline 
\end{tabular}
\tablefoot{\tablefoottext{a}{Measured temperature with an estimated error of 250 K \citep{Brown2011,Huber2014}.}}
\end{table}

\subsection{The case of the outsiders}
\label{ss:out}

We found that only a 2\% of the \dScu stars of our sample lie outside the limits of the predicted dispersion due to the gravity-darkening effect. Most of those stars could be included if we take into account their 2- or 3-$\sigma$ error bars (see Fig.~\ref{f:Turel}). Nevertheless, there are many possible physical reasons that could cause a higher departure from the effective temperature of a spherically symmetric star.\\

One of our hypotheses was that the higher limit of the rotation rate for a stable star is Eq.~\ref{e:OmegaK}. Considering the possibility of finding stars with a higher rotation rate, we took into account a new higher limit,
\begin{equation}
\Omega \approx \Omega_{C}=\sqrt{g_{\rm eff,e} / R_{e}} \simeq 1.18 \Omega_{K}\; ,
\label{e:OmegaC}
\end{equation}
where $g_{\rm eff,e}$ is the effective gravity, and
\begin{equation}
R_{e} = R \sqrt{1+\epsilon^{2}/3} \; ,
\label{e:Re}
\end{equation}
the radius both at the equator assuming a constant volume compared with a spherically symmetric star. Therefore, the proportion of \dScu stars inside the predicted dispersion increases to 99.3\% (see Table~\ref{t:out}).\\

\subsection{The case of the HADS}
\label{ss:hads}

\begin{table*}
\caption{Rotation rate and inclination limits of the HADS candidates in our star sample.}
\label{t:hads}
\centering
\begin{tabular}{c c c c c c c c}
KIC & $A_{0}$ (ppm)& $\nu_{\rm max}$ ($\mu$Hz)& $\bar{T}_{\rm eff}$ (K)& $T_{\rm eff}$ (K)\tablefootmark{a} & $\frac{\Omega}{\Omega_{K}}$ (\%) & $i$ ($^{\circ}$)\\
\hline
2857323&  91200 $\pm$ 500&  180 $\pm$    30& 7520 $\pm$   100& 7960&$\gtrsim$  81& $\lesssim$ 31\\
5950759& 122700 $\pm$ 600&  210 $\pm$    30& 7600 $\pm$   100& 8040&$\gtrsim$  82& $\lesssim$ 31\\
9408694& 158400 $\pm$ 800&   90 $\pm$    20& 7260 $\pm$    70& 6810&$\gtrsim$  86& $\gtrsim$ 76\\
\hline 
\end{tabular}
\tablefoot{\tablefoottext{a}{Measured temperature with an estimated error of 250 K \citep{Brown2011,Huber2014}.}}
\end{table*}  

The High Amplitude \dScu stars should be good candidates for testing this scaling relation. These are \dScu stars with amplitudes around or higher than $10^5$ ppm with temperatures between 7000 and 8000 K, approximately \citep{McNamara2000}. \cite{Breger2000} points out that most stars of this kind have a slow projected velocity. In this sense, HADS seem to rotate slowly or to have an inclination close to pole-on. As a consequence, we should find them close to the $i-\Omega$ degeneracy zone or also with higher temperatures (see KIC~2857323 and KIC~5950759 in Table~\ref{t:hads}).\\

In contrast, \cite{Balona2012} find that KIC~9408694 has an uncommon higher projected velocity $v \sin i \sim 100$ km/s. Previous studies find that only 3 of 22 have $v \sin i > 40$ km/s \citep{Rodriguez2000}. \cite{Balona2012} also estimate that the rotational rate is lower than  $\Omega \lesssim 15 \pm 4 $ $\mu$Hz taking into account that the mass and the radius of the star is $M\sim 2.2 \pm 0.2 M_{\odot}$, and $R \sim 4.5 \pm 0.7 R_{\odot}$, respectively. The lowest frequency peak observed is $\nu \sim 34$ $\mu$Hz and the authors discard this as a rotational signal. However, we noted that half of the observed signal ($\sim 17$ $\mu$Hz) is within the limits. Using the difference of temperature due to the gravity-darkening effect, we delimited the rotation rate to $\Omega \gtrsim 0.86\Omega_{K}$ and the inclination to $i \gtrsim 76^{\circ}$ (see Table~\ref{t:hads}). Therefore, KIC~9408694 might be an equator-on HADS showing twice the rotational frequency approximately.\\

We find that all three HADS of our sample are extreme rotators with low inclination angle with the exception of the studied special case of  KIC~9408694. Therefore, our scaling relation seems also useful to characterize this kind of stars. The low incidence of HADS in our sample, $\sim 0.28$\% is in agreement with the observations of other authors \citep{Lee2008,Bowman2016}.

\section{Conclusions}
\label{s:conclusion}

The oscillation power spectra of 700 \dScu stars observed by \textit{Kepler} have been obtained and their overall envelope parametrized. These envelopes could be highly asymmetric compared with low mass stars, main sequence stars, and red giants, especially for those \dScu stars with asymmetry towards lower frequencies. This asymmetry may be produced by an asymmetry in the energy balance responsible of the excitation of the modes.\\

Furthermore, we find that the frequency at maximum power is linearly related with the effective temperature. This could be explained by the excitation of higher radial order modes in \dScu type stars with higher temperatures as predicted by \citet{Dziembowski1997}. This is an important result because knowing a power spectral characteristic we will directly obtain an intrinsic stellar parameter, independent of rotation and the point of view of the observer. The scatter of points around the fitted straight line can be explained as the gravity-darkening effect due to the oblateness of the star and its inclination with respect the line of sight. This property allows us to delimit both rotation and inclination from the line of sight, especially for extreme rotators. The well-known \dScu stars observed by \textit{CoRoT} sismo-channel gives support to our findings.\\

Our next step is to use these delimited magnitudes to make an improved analysis of the regularities to find the rotation and the large separation. In this sense, we will be able to characterize \dScu with a few seismic indices ($\Omega$, $\Delta \nu$, $\nu_{\rm max}$) as it has been done to solar-type pulsators.\\

\paragraph{}
\begin{acknowledgements}
Comments from A. Moya are gratefully acknowledged. The authors wish to thank the referee for useful suggestions that improved the paper. We also thank the \textit{CoRoT} and \textit{Kepler} Teams whose efforts made these results possible. The \textit{CoRoT} space mission has been developed and was operated by \textit{CNES}, with contributions from Austria, Belgium, Brazil, ESA (RSSD and Science Program), Germany and Spain. Funding for \textit{Kepler}'s Discovery mission is provided by NASA’s Science Mission Directorate. S.B.F. wishes to thank the Solar Physics team of the \textit{Universitat de les Illes Balears} (UIB) for their hospitality during his stay in Majorca. He has received financial support from the grants AYA2010-20982-C02-02, ESP2015-65712-C5-1-R, and ESP2017-87676-C5-1-R. R.A.G. acknowledges the financial support from the ANR (Agence Nationale de la Recherche, France) program IDEE (n ANR-12-BS05-0008) “Interaction Des Étoiles et des Exoplanètes” and from the CNES GOLF and PLATO grants at CEA.
\end{acknowledgements}

\bibliographystyle{aa}
\bibliography{tcb}

\end{document}